\documentclass[journal=nalefd, manuscript=article]{achemso}
\setkeys{acs}{articletitle=true}

\usepackage[version=3]{mhchem} 
\usepackage{amsmath}
\usepackage[colorlinks = true,
linkcolor = blue,
urlcolor = blue,
citecolor = blue,
anchorcolor = blue]{hyperref}
\usepackage{pifont}
\usepackage{ulem}

\newcommand{\WSeTe}{WSe\textsubscript{2-2x}Te\textsubscript{2x}}
\newcommand{\dose}{e\textsuperscript{-}/nm\textsuperscript{2}}
\newcommand{\Ang}{\text{\AA} }

\author{Chia-Hao Lee}
\affiliation{Department of Materials Science and Engineering, University of Illinois Urbana-Champaign, Urbana, IL, United States 61801}

\author{Abid Khan}
\affiliation{Department of Physics, University of Illinois Urbana-Champaign, Urbana, IL, United States 61801}
\altaffiliation{These authors contributed equally to this work}

\author{Di Luo}
\affiliation{Department of Physics, University of Illinois Urbana-Champaign, Urbana, IL, United States 61801}
\altaffiliation{These authors contributed equally to this work}

\author{Tatiane P. Santos}
\affiliation{Department of Materials Science and Engineering, University of Illinois Urbana-Champaign, Urbana, IL, United States 61801}

\author{Chuqiao Shi}
\affiliation{Department of Materials Science and Engineering, University of Illinois Urbana-Champaign, Urbana, IL, United States 61801}

\author{Blanka E. Janicek}
\affiliation{Department of Materials Science and Engineering, University of Illinois Urbana-Champaign, Urbana, IL, United States 61801}

\author{Sangmin Kang}
\affiliation{Department of Electrical and Computer Engineering, University of Illinois Urbana-Champaign, Urbana, IL, United States 61801}

\author{Wenjuan Zhu}
\affiliation{Department of Electrical and Computer Engineering, University of Illinois Urbana-Champaign, Urbana, IL, United States 61801}

\author{Nahil A. Sobh}
\affiliation{Beckman Institute for Advanced Science and Technology, University of Illinois Urbana-Champaign, Urbana, IL, United States 61801}

\author{Andr\'e Schleife}
\affiliation{Department of Materials Science and Engineering, University of Illinois Urbana-Champaign, Urbana, IL, United States 61801}
\alsoaffiliation{Materials Research Laboratory, University of Illinois at Urbana-Champaign, Urbana, IL 61801, USA}
\alsoaffiliation{National Center for Supercomputing Applications, University of Illinois at Urbana-Champaign, Urbana, IL 61801, USA}

\author{Bryan K. Clark}
\affiliation{Department of Physics, University of Illinois Urbana-Champaign, Urbana, IL, United States 61801}

\author{Pinshane Y. Huang}
\affiliation{Department of Materials Science and Engineering, University of Illinois Urbana-Champaign, Urbana, IL, United States 61801}
\alsoaffiliation{Materials Research Laboratory, University of Illinois at Urbana-Champaign, Urbana, IL 61801, USA}

\email{pyhuang@illinois.edu}

\title{Deep Learning Enabled Strain Mapping of Single-Atom Defects in 2D Transition Metal Dichalcogenides with Sub-picometer Precision}

\begin{document}

\begin{abstract}
2D materials offer an ideal platform to study the strain fields induced by individual atomic defects, yet challenges associated with radiation damage have so-far limited electron microscopy methods to probe these atomic-scale strain fields. Here, we demonstrate an approach to probe single-atom defects with sub-picometer precision in a monolayer 2D transition metal dichalcogenide, \WSeTe. We utilize deep learning to mine large datasets of aberration-corrected scanning transmission electron microscopy images to locate and classify point defects. By combining hundreds of images of nominally identical defects, we generate high signal-to-noise class-averages which allow us to measure 2D atomic coordinates with up to 0.3 pm precision. Our methods reveal that Se vacancies introduce complex, oscillating strain fields in the \WSeTe lattice which cannot be explained by continuum elastic theory. These results indicate the potential impact of computer vision for the development of high-precision electron microscopy methods for beam-sensitive materials.
\end{abstract}

\subsection{Keywords}
Deep learning, fully convolutional network (FCN), single-atom defects, strain mapping, scanning transmission electron microscopy, 2D materials
\newline
\\A key challenge in characterizing 2D materials is determining the structure of defects with picometer precision. Defect and strain engineering of 2D materials are emerging tools to tune the optical and electronic properties of atomically thin layers\cite{Feng2012, VanDerZande2013, Lin2016}. Yet, while techniques such as aberration-corrected scanning transmission electron microscopy (STEM) have the ability to image each atom in 2D materials, the precision of atom-by-atom electron microscopy has so far been limited to the scale of 8-20 picometers, or strains on the order of 3\% or more\cite{Huang2013a, Azizi2017, Wang2016a}. While these methods can detect the relatively large strains at the nearest-neighbor sites of vacancies, the local strains ($\approx$ 1\%) expected to result from substitutions and long-range strain fields from point defects have so far been below the detection limits of atomic-resolution (S)TEM.

This precision is fundamentally limited by radiation damage from the electron beam: high radiation doses are required to precisely measure the position of single atoms, yet ionization and knock-on damage alter the structure of defects at high electron dose\cite{Komsa2013, Algara-Siller2013, Elibol2018}. In bulk materials where the precision is limited by microscope instabilities rather than electron beam damage, the measurement precision can be enhanced by acquiring a series of images on the same region, then combining the resulting data using techniques such as drift correction (10 pm)\cite{Ophus2016}, template matching (5-15 pm)\cite{Mevenkamp2015, Zuo2014}, rigid (5 pm)\cite{Kimoto2010} and non-rigid registration (0.3-0.9 pm)\cite{Yankovich2014}. On their own, these approaches have limited utility for measuring the intrinsic structure of 2D materials because they typically require high doses on the order of $10^8 - 10^9$ \dose, above the damage thresholds for many 2D materials. For example, serious electron beam damage of free-standing, monolayer \ce{MoS2} has been observed after an electron dose of $2.8 \times 10^8$ \dose at 80 kV\cite{Algara-Siller2013}. Meanwhile, diffraction-based strain measurements such as nanobeam electron diffraction \cite{Han2018} can measure subpicometer strains in 2D TMDCs, but are limited to a spatial resolution of a few nanometers. These challenges mean that for 2D materials, existing techniques exhibit a trade-off between spatial resolution and the precision with which strain can be measured, making it difficult to measure the strain field of atomic defects. Yet at the same time, 2D materials offer a profound opportunity for understanding atomic-scale strain. Because they are only a single unit cell thick, 2D materials are ideal for demonstrating high precision characterization methods, such as the ability to characterize how each atom in a material responds to local perturbations\cite{Huang2013a, Warner2012, Azizi2017}. 

Here, we apply machine learning to locate and classify each point defect in large datasets of atomic-resolution images, then use the resulting data to generate class-averaged images of single-atom defects in 2D materials. This method enables sub-picometer precision measurements of beam-sensitive structures because it combines information measured from large numbers of nominally identical defects while limiting the dose to any individual atom. Our approach is analogous to the class-averaging methods used in single particle cryo-electron microscopy, where they are used to aid in solving the structure of biological macromolecules and viruses\cite{Cheng2009}.

We demonstrate our approach using an alloyed 2D transition metal dichalcogenide (TMDC), monolayer \WSeTe. Previously, STEM has been used to directly measure the local variations in the concentration, ordering, and properties of alloyed TMDCs. \cite{Apte2018, Lin2018, Tizei2015, Rhodes2017}. We synthesized 2H-\WSeTe using cooling-mediated, one-step chemical vapor deposition (CVD) on \ce{SiO2}/\ce{Si} substrates. The \WSeTe was then transferred to TEM grids using a wet-transfer technique (see Supporting Information (SI)). These methods produce suspended flakes of predominately monolayer \WSeTe that are 10 - 20 $\mu$m across. These \WSeTe samples naturally contain point defects including Te substitutions and Se vacancies which provide local lattice distortions that can be used to test our techniques.

\begin{figure}[H]
\includegraphics[width= 16 cm]{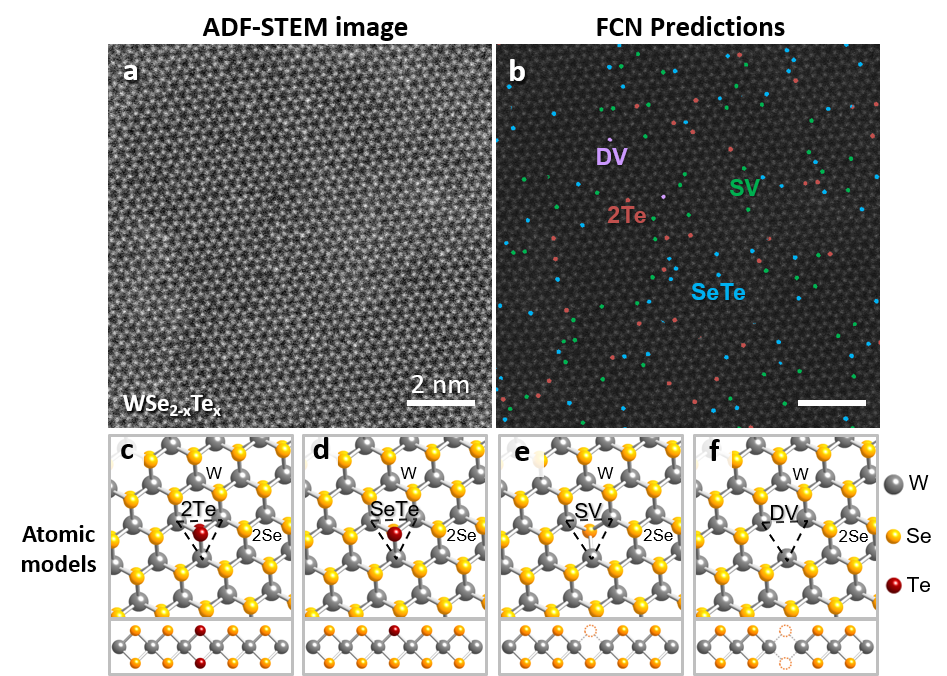}
\caption{Deep learning-enabled identification and classification of point defects in ADF-STEM image. (a) Atomic-resolution ADF-STEM image of \WSeTe. (b) Chalcogen-site defects identified by fully convolutional networks (FCNs) overlaid on image from (a). Labels indicate one or two Te substitutions ($SeTe$ and $2Te$, respectively) and single or double Se vacancies ($SV$ and $DV$). (c-f) Top and side-view schematics of defect structures. The chalcogen defect centers are marked with dashed triangles.
}
\label{fgr:1}
\end{figure}

We next acquired aberration-corrected annular dark-field (ADF) STEM images (\ref{fgr:1}a) and used machine learning to locate and classify the defects present, as illustrated in Figure \ref{fgr:1}. For this study, we analyzed images of 9 different regions, spanning a total area of 4000 nm\textsuperscript{2}, or approximately 130,000 atoms. To analyze the data, we trained a deep learning model based on fully convolutional networks (FCNs) with ResUNet architecture to locate and classify the point defects in \WSeTe, producing 2D maps of the defect positions (\ref{fgr:1}b). Neural networks have already revolutionized image recognition in fields such as medical diagnosis, weather forecasting, and facial recognition; recently, they have also been applied to identify atomic defects in atomic-resolution (S)TEM images\cite{Madsen2018, Ziatdinov2017a, Maksov2019}. Conventionally, defect detection has been a labor-intensive task which is often done by hand\cite{Rhodes2017, Azizi2017} or simple image processing such as Fourier filtering\cite{Lin2013} or intensity thresholding\cite{Gong2014, Lin2018}. Neural networks offer an opportunity to automate defect identification, making it possible to efficiently locate large numbers of defects to generate class averages systematically while minimizing human intervention. 
We trained FCNs using simulated data generated via incoherent image simulations using Computem\cite{Kirkland2013a, Kirkland2013}. In order to make our simulations more realistic, we apply a set of post-processing steps to the images, including the addition of Gaussian noise, probe jittering, image shear, and varying spatial sampling, to create our final training data. Similar methods are well-established in the literature \cite{Ziatdinov2017a, Maksov2019, Madsen2018}, though we found that we achieved the highest classification precision on experimental data by introducing low-frequency contrast variations in the simulated data to emulate surface contamination. We found that these methods yielded a true positive rate of 98\%. When we compared the true positive rate with FCNs trained directly on hand-labeled experimental data, we found that the simulation-trained data performed comparably to FCNs trained on experimental data but with considerably less manual labor (see SI for evaluation metrics). The source codes for training set generation and model training are freely available on \textbf{Github} at : \href{http://github.com/ClarkResearchGroup/stem-learning/}{github.com/ClarkResearchGroup/stem-learning/}

We focused on the four primary types of chalcogen-site defects present in our samples, which we refer to as $2Te$, $SeTe$, $SV$, and $DV$ (see Figure \ref{fgr:1}c-f). Our naming convention describes the composition and filling of the chalcogen sites in \WSeTe. In projection, the chalcogen columns can contain either two Se atoms (no defects, $2Se$), one or two Te substitutions ($SeTe$ and $2Te$, respectively), or one or two Se vacancies ($SV$ and $DV$). These defects are the most common point defects we observed in \WSeTe. Using the large datasets probed by FCN, we conducted population analyses of the defects present in \WSeTe. We calculated both the total number and concentrations (over all 86000 chalcogen sites) of each defect type in our experimental images. We found the Te fraction in our samples is \WSeTe where x = 0.06. Meanwhile, $3\%$ of chalcogen sites are occupied by vacancies; this number is an upper bound of the as-grown vacancy concentration because TEM sample fabrication and electron irradiation can induce additional vacancies. We found that metal-site defects were extremely rare (comprising less than 0.04\% of metal sites), and we did not observe columns containing a single Te atom ($1Te$).

\begin{figure}[H]
\includegraphics[width= 16 cm]{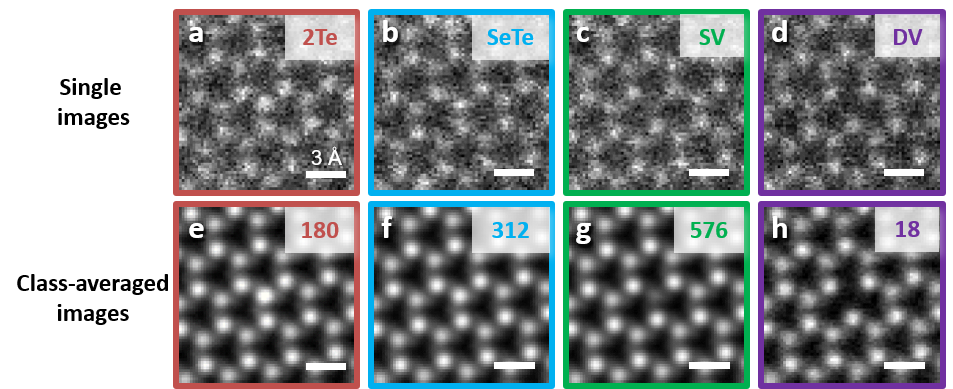}
\caption{Comparisons between single and class-averaged images of $2Te$, $SeTe$, $SV$, and $DV$ defects. (a-d) Representative single images of FCN-identified defects sectioned from Figure \ref{fgr:1}(a). By aligning and summing many equivalent lattice sites using rigid-registration, we produce high SNR, class-averaged images (e-h) from nominally identical point defects. The number of images summed is labeled at the top right corner of each image.
}
\label{fgr:2}
\end{figure}

Next, we generated class-averaged images of each defect type from the FCN outputs (Figure \ref{fgr:2}). From the thousands of defects identified via the FCN, we selected only isolated defects -- i.e. defects that were separated by a distance $d\geq 6.6$ \Ang (roughly $4 \times 4$ unit cells) from any other defects. This step dramatically reduced the number of defects used for class-averaging, but it allowed us to study the structure of the defects with minimal external perturbations. The use of FCNs enabled this step because it allowed us to locate a sufficiently large population to retain several hundred defects in each class after this filtering step. We then sectioned the original images into small windows centered around each individual defect as shown in Figure \ref{fgr:2}a-d. The sectioned images were grouped by defect type, creating image stacks containing 180 $2Te$, 312 $SeTe$, 576 $SV$, 18 $DV$ individual defects, and 437 defect-free $2Se$ regions. Finally, we aligned and summed each image stack using rigid registration\cite{Savitzky2018}, producing the high signal-to-noise ratio (SNR) class-averaged images shown in Figure \ref{fgr:2}e-h. 

\begin{figure}[H]
\includegraphics[width= 16 cm]{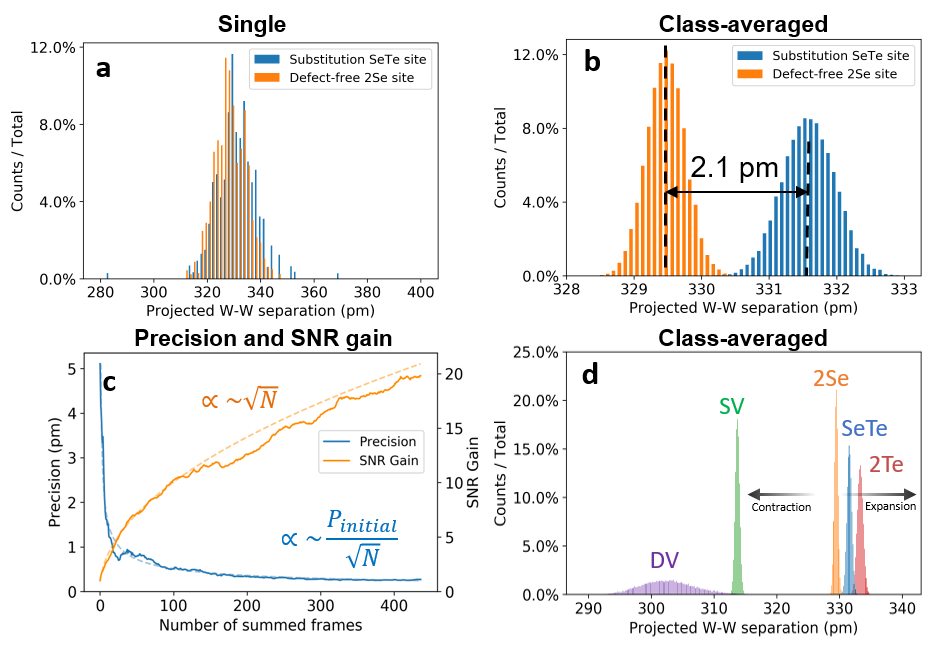}
\caption{Impact of class-averaging on signal-to-noise and precision of atomic separations. (a-b) Distributions of projected W-W separations nearest to the defect site in (a) individual (b) class-averaged images generated by summing 312 single Te substitution ($SeTe$) and 437 defect-free ($2Se$) images respectively. Class-averaged distributions are generated through bootstrapping. Unlike in the individual images, the class-averaged images show well-separated distributions of W-W separation measurements of $SeTe$ substituted and defect-free $2Se$ sites. From class averaging, the measured W-W separation is $331.6 \pm 0.4$ pm at $SeTe$ substituted sites and $329.5 \pm 0.3$ pm at defect-free $2Se$ sites. (c) Precision and SNR gain as a function of summed frames $N$. The precision scales with $P_{initial}/\sqrt{N}$, while the SNR gain scales with $\sqrt{N}$ due to the reduction of Poisson noise. (d) Distributions of projected W-W separations measured on class-averaged images of defect-free $2Se$ sites and each defect types ($2Te$, $SeTe$, $SV$, $DV$), which yielded local strain of $1.2 \pm 0.2 \%$, $0.6 \pm 0.2 \%$, $-4.8 \pm 0.1\%$, and $-8 \pm 1\%$ respectively.
}
\label{fgr:3}
\end{figure}

As shown in Figure \ref{fgr:3}, class averaging enables sub-picometer precision measurements of atomic coordinates and local strains. In Figure \ref{fgr:3}a, we used 2D Gaussian fitting to determine the positions of atomic columns in a series of single images, measure the 3 nearest W-W spacings around $SeTe$ substitutions, and compare them with the same measurements in defect-free images. We obtained W-W spacings of $330 \pm 8$ pm (std. dev.) for the $SeTe$ substitution and $330 \pm 6$ pm for defect-free $2Se$ sites. The histograms in Figure \ref{fgr:3}a overlap heavily, indicating that single images cannot be used to distinguish the local strains around a single Te substitution. 

In contrast, Figure \ref{fgr:3}b shows the well-separated distributions of W-W spacings measured from class-averaged images. To generate these distributions, we apply a bootstrap approach commonly used in statistical analysis\cite{Efron1979} to produce several class-averaged images using randomly selected subsets of images from the original image stack. These bootstrapped class averages allow us to estimate the measurement precision using the same definition as for single images (see SI). For the class-averaged data, we measure W-W separations of $331.6 \pm 0.4$ pm around the Te substitution (summing 312 images for each class average), and $329.5 \pm 0.3$ pm for the defect-free site (using 437 images). These data show the utility of class averaging, which provided a 21-fold improvement in precision when summing 437 images, sufficient to measure local strains on the order of 0.1 $\%$. Notably, the sub-pm precision obtained using our class-averaging approach is comparable to the highest precision electron microscopy measurements obtained via multi-frame averaging in bulk materials\cite{Yankovich2014}, but without increasing the dose per unit area. This approach allows us to access the small strains around atomic defects in 2D materials while minimizing electron beam damage.

The sub-pm precision obtained in the class-averaged images is a direct result of their increased SNR. Figure \ref{fgr:3}c plots the precision and SNR gain in class-averaged images as a function of the number of images summed for defect-free regions. The gain in the SNR, which is defined as $SNR_{sum}/SNR_{raw}$, scales as ($\sqrt{N}$), where $N$ is the number of images summed (see SI). Meanwhile, the measurement precision of the atomic spacings scales as $P_{initial}/\sqrt{N}$, or proportional to the inverse of the SNR. These scaling laws arise because Poisson noise is the dominant source of noise in the ADF-STEM detector\cite{VanAert2002,Savitzky2018}. Figure \ref{fgr:3}d shows the distributions of nearest neighbor W-W atomic spacings for each defect type after class-averaging. The variation in the widths of these distributions, such as the wide distribution of $DV$, mainly results from differences in the number of defects summed. The distributions for each defect type are well separated, indicating that we are able to distinguish the local lattice expansion from single and double Te substitutions as well as the contraction that results from single and double Se vacancies. 

\begin{figure}[H]
\includegraphics[width= 16 cm]{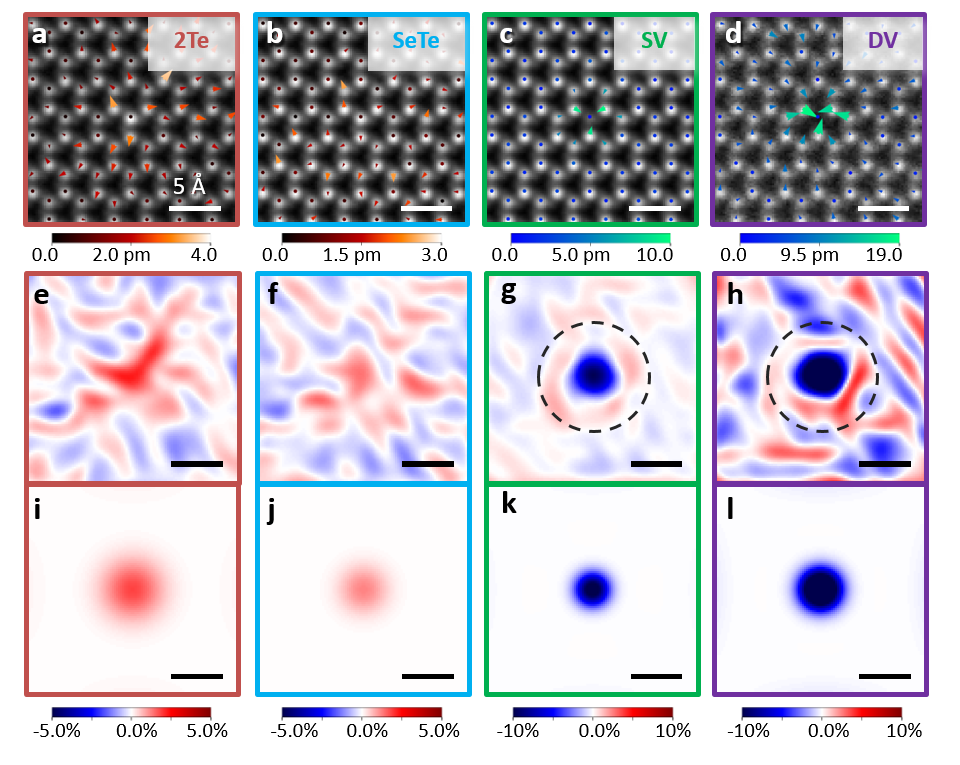}
\caption{Displacement and strain fields for chalcogen site defects. (a-d) Two-dimensional displacement vector field overlaid on class-averaged images of chalcogen site defects. The vectors are enlarged for visibility by 40 times in (a-b) and 10 times in (c-d). (e-h) Experimental dilation fields calculated from the displacement fields. The dilation corresponds to the local projected area change. (e) $2Te$ and (f) $SeTe$ exhibit local expansion, while (g) $SV$ and (h) $DV$ exhibit local contraction. (i-l) Best-fit dilation fields calculated with 2D isotropic elastic continuum theory using Eshelby's inclusion model. 
}
\label{fgr:4}
\end{figure}

Next, we measured the displacement and strain fields for each defect type. Figure \ref{fgr:4}a-d shows magnified 2D displacement vectors overlaid on class-averaged defect images. Displacement vectors are obtained by comparing the positions of each atomic column on class-averaged defect images to the positions measured in a defect-free, class-averaged reference image. Single ($SV$) and di-vacancies ($DV$) correspond with a local contraction of the lattice, while single ($SeTe$) and double ($2Te$) substitutions produce a local expansion. The magnitude of the displacement vectors decays quickly as a function of distance from the defect centers, for example dropping below 1 pm within 3 unit cells for a single Te vacancy. To better visualize the local distortions, we calculate the 2D strain tensor components $\epsilon_{xx}, \epsilon_{xy}$, and $\epsilon_{yy}$, for each defect type from their displacement vectors (see Figure S7 for calculations and all strain tensor components of $SV$). Density functional theory (DFT) simulations indicate that these in-plane strain components are much larger than out-of-plane deformation, and that as a result the 2D strain fields measured from the 2D projections in STEM images are a good approximation of the full 3D deformation. 

Figures \ref{fgr:4}e-h show the experimental dilation maps, which correspond to local 2D area change associated with each defect, calculated as the sum of the diagonal components $\epsilon_{xx}+\epsilon_{yy}$ of the strain tensor. We compared these experimental dilation fields to those calculated using a purely elastic continuum theory. To calculate the strain field that would result from an ideal elastic medium, we use the 2D version of Eshelby's inclusion model, where the crystal is modeled as an infinite, isotropic 2D elastic continuum under deformation from a point-like inclusion \cite{Eshelby1957, Mura1982, Kolesnikova2014}. While the best-fit elastic models (Figure \ref{fgr:4}i-l) capture the behavior of the experimental dilation fields near the defect core (Figure \ref{fgr:4}e-h), we also notice key differences, as discussed below. 

\begin{figure}[H]
\includegraphics[width= 13.3 cm]{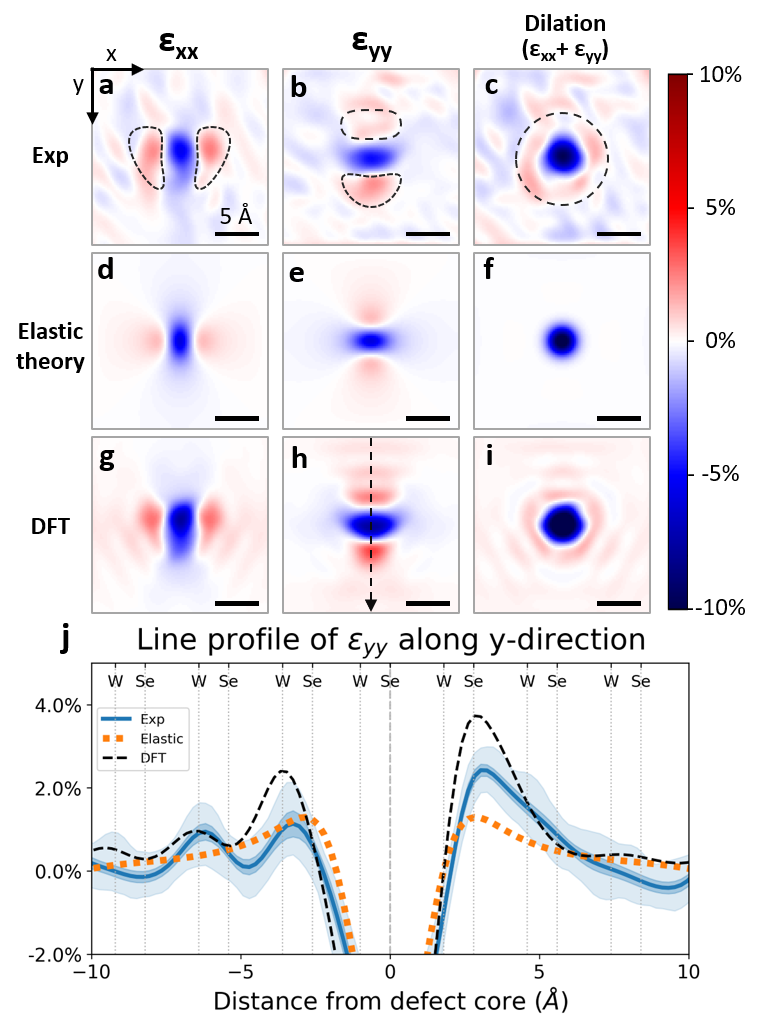}
\caption{Strain fields at single Se vacancy. (a-c) Experimental strain fields calculated from the derivative of displacement field in Figure \ref{fgr:4}(c). (d-f) Best-fit strain fields calculated by elastic theory using Eshelby's inclusion model. (g-i) Strain fields calculated from DFT simulations of defect relaxation. The anisotropic features show good match with the experimental data (a-b). (j) Line profiles of experimental, elastic theory, and DFT-derived $\epsilon_{yy}$ across the vacancy, as marked by dashed arrow. The shaded regions of the experimental line profile correspond respectively to $\pm 1$ standard deviation ($\pm 0.2\%$) and the full-range of the experimental distribution of strain values measured using bootstrapping. In contrast to the monotonically decaying strain field predicted by continuum elastic theory, both experimental and DFT profiles show oscillations in the strain field. Vertical lines indicate the locations of W and Se columns.
}
\label{fgr:5}
\end{figure}

Figure \ref{fgr:5} compares the $\epsilon_{xx}$, $\epsilon_{yy}$, and dilation components of experimental strain fields from a single vacancy (Figure \ref{fgr:5}a-c) to both a best-fit elastic continuum theory (Figure \ref{fgr:5}d-f), and first-principles simulations using density functional theory (DFT) (Figure \ref{fgr:5}g-i). In order to calculate strain fields from DFT simulations, we conducted structural relaxations of both the $SV$ and defect-free lattice (see SI). A 9x9 supercell was required to avoid coupling of the strain field between defects when using periodic boundary conditions. Using the relaxed atomic coordinates, we simulated ADF-STEM images using a multislice algorithm implemented in Computem \cite{Kirkland2013a} and then applied the same methods used for our experimental data to calculate the DFT-derived strain fields. As shown in Figure \ref{fgr:5}, our increased experimental precision allows us to observe fine features in the strain field that deviate clearly from continuum elastic theory but are in good agreement with DFT. The experimental strain fields deviate from continuum elastic theory in two main ways. First, they are not isotropic in 2D but instead reflect the symmetry of the lattice. For example, the regions marked by the black dashed lines in Figure \ref{fgr:5}a have higher intensity in the top half of the image. Similarly, the experimental $\epsilon_{yy}$ is asymmetric across the center of the defect (Figure \ref{fgr:5}b). These asymmetries are present in the DFT simulations (Figure \ref{fgr:5}g-h), but not in the continuum elastic model, which predicts two-fold symmetric $\epsilon_{xx}$ and $\epsilon_{yy}$ (Figure \ref{fgr:5}d-e). Second, while the dilation field calculated using the continuum elastic model indicates only local contraction around the vacancy (Figure \ref{fgr:5}f), we observe both contraction and expansion in the DFT data( \ref{fgr:5}i) and in experiment (dashed circle in Figure \ref{fgr:5}c). In other words, the nearest unit cells expand while the defect core strongly contracts, forming an oscillating strain field. To our knowledge, these defect-induced strain field oscillations have never been experimentally observed in 2D materials. 

We investigate this phenomenon in more detail in the line profile of $\epsilon_{yy}$ in Figure \ref{fgr:5}j. In this plot, the mean experimental strain profile is shown in dark blue, while blue shading indicates the range of strain values calculated via bootstrapping. In the continuum elastic model (orange), the strain field monotonically decays away from the defect core, whereas the DFT (black) and experiment (blue) show clear oscillations up to a nanometer away from the defect core. Overall, we find excellent agreement between experiment and DFT-PBE, particularly for the locations of maxima and minima of strain field oscillations. We do note that the peak experimental strain field is smaller in magnitude than in the DFT. This likely occurs because DFT-PBE is known to underestimate elastic constants relative to experiment with a generalized-gradient approximation for exchange and correlation \cite{Rasander2015}. Passivation of some vacancies is another possible contributing factor.

To understand the origin and significance of these observed strain field oscillations, we note that similar phenomena have been predicted in both bulk metals\cite{Girifalco1960} and ceramics\cite{Hardy1960}. In metals, oscillating strain fields may arise from defect-induced charge redistribution such as Friedel oscillations\cite{Singhal1973}, while for ionic crystals, Coulomb interactions between charge perturbations at defect site and ion cores of opposite signs lead directly to oscillations in the strain field. Either of these effects may contribute to the features we see in \WSeTe. Accurate models for these complex strain fields were part of the historical motivation for the development of lattice static methods such as the Kanzaki method\cite{Kanzaki1957} and Green's function methods for modeling point defects in crystals\cite{Tewary1973, Trinkle2008}. In this context, direct observation of oscillating strain fields in 2D materials indicates both a new milestone in the ability to test and refine high-accuracy mechanical models for defects in crystals and a need to account for long-range strain fields when modeling defects in atomically thin materials.

In conclusion, we have developed techniques based on machine learning and aberration-corrected STEM to visualize the strain fields induced by single-atom defects in 2D materials. We used these methods to directly observe the strain fields of vacancies and substitutions in \WSeTe, where the sub-pm precision enabled by class-averaging revealed oscillations in the strain field around chalcogen vacancies that deviate from isotropic elastic continuum theory but agree well with DFT simulations. A key advantage of these methods is that they enable high precision measurements of beam-sensitive materials by leveraging computer vision to mine atomic-resolution datasets without requiring any changes in instrumentation. These methods should be particularly useful for studying 2D materials and other radiation-sensitive crystals. Going forward, our deep learning enabled class-averaging can be applied in principle to any atomic resolution electron microscopy datasets, including spectrum imaging and 4D STEM.

\subsection{\textbf{\ding{110} ASSOCIATED CONTENT}}
\textbf{Supporting Information}\\ 
\WSeTe synthesis, TEM sample fabrication, ADF-STEM acquisition parameters, FCN model architecture and training setup, FCN model performance evaluation, bootstrapping process and strain analysis, and details of DFT calculation.

\subsection{\textbf{\ding{110} AUTHOR INFORMATION}}
\textbf{Corresponding Author}\\
*Email: \href{pyhuang@illinois.edu}{pyhuang@illinois.edu}\\
\textbf{ORCID}\\
Chia-Hao Lee: 0000-0001-8567-5637\\
Abid Khan: 0000-0002-0450-0729\\
Di Luo: 0000-0001-6562-1762\\
Blanka E. Janicek: 0000-0002-5529-2819\\
Pinshane Y. Huang: 0000-0002-1095-1833\\
Andr\'e Schleife: 0000-0003-0496-8214\\
\textbf{Author Contributions}\\
Under supervision by P.Y.H., C.-H.L. analyzed the preliminary STEM images acquired by B.E.J. and performed \WSeTe TEM sample preparation, STEM imaging, data analysis, and elastic theory modeling. Under supervision by P.Y.H., C.-H.L. and C.S. generated simulated STEM images for FCNs training. Under supervision by B.K.C., D.L. and A.K. constructed the FCN models. N.A.S. contributed to the FCN structure and performance evaluation. Under supervision by A.S., T.S. performed DFT calculations. Under supervision by W.Z., S.K. synthesized the 2D \WSeTe flakes. All authors read and contributed to the manuscript.\\
\textbf{Notes}\\
The authors declare no competing financial interest.

\begin{acknowledgement}
This work was primarily funded by the Air Force Office of Scientific Research under Award Number FA9550-17-1-0213 and the U. S. Department of Energy, Office of Science, Office of Basic Energy Sciences, under Award Number DE-SC0020190. Zhu and Kang acknowledge support from the Office of Naval Research under Award Number NAVY N00014-17-1-2973. Schleife and Santos acknowledge support from the Office of Naval Research (grant No.\ N00014-18-1-2605). This research is part of the Blue Waters sustained-petascale computing project, which is supported by the National Science Foundation (awards OCI-0725070 and ACI-1238993) and the state of Illinois. Blue Waters is a joint effort of the University of Illinois at Urbana-Champaign and its National Center for Supercomputing Applications. This work also made use of the Illinois Campus Cluster, a computing resource that is operated by the Illinois Campus Cluster Program (ICCP) in conjunction with the National Center for Supercomputing Applications (NCSA) and which is supported by funds from the University of Illinois at Urbana-Champaign. This work was carried out in part in Materials Research Laboratory Central Facilities at the University of Illinois.

The authors thank Prof. Paul Voyles, Prof. Elif Ertekin, Dr. Colin Ophus, and Prof. Dallas Trinkle for helpful discussions.

\end{acknowledgement}

\bibliography{Manuscript}

\newpage
\subsection{Graphical TOC Entry}
\begin{figure}[H]
\includegraphics[width= 8.2 cm]{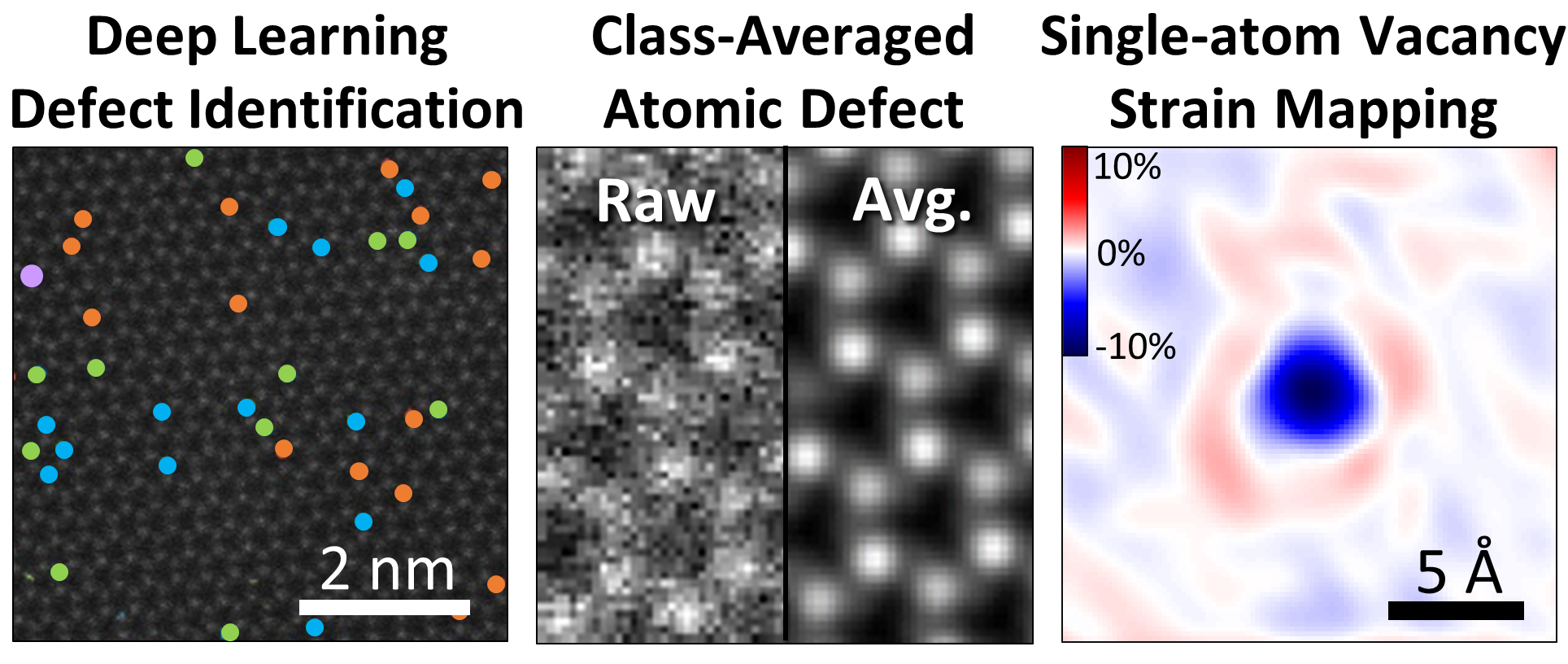}
\label{TOC}
\end{figure}

\end{document}


\section{1. STEM experiment}
\subsection{1.1 \WSeTe synthesis and TEM sample fabrication}
monolayer \WSeTe was synthesized by cooling-mediated chemical vapor deposition with a tube furnace system. High purity Se powder, Te powder, and \ce{WO3} powder were used as precursors (Sigma Aldrich, $> 99.9 \%$) with high purity Ar/\ce{H2} as the carrier gas. Se powder and Te powder were placed upstream at the edge of the heating zone, while \ce{WO3} powder was placed near the center of the heating zone. \ce{SiO2} (280 nm)/Si wafers were placed on top of the \ce{WO3} crucible located near the end of the heating zone. The growth phase was set to $760^\circ$C for 10 min. After the synthesis, the samples were cooled down to room temperature with a cooling rate $ > {100}^\circ$C/min using water-assisted cooling. In order to prepare the samples for STEM analysis, the as-synthesized \WSeTe\ flakes were first coated with a polymethyl methacrylate (A4 PMMA) supporting layer, then the \ce{SiO2}/Si substrate was removed after soaking in 5M \ce{KOH} solution for 1 hr. The PMMA/\WSeTe\ stack was rinsed with a series of water baths, then scooped onto a holey carbon TEM grid (Au Quantifoil\textsuperscript{\textregistered}) with 2 $\mu$m holes. The PMMA film was then removed by series of solvent baths using acetone and isopropanol (IPA), exposing the suspended \WSeTe\ flakes on TEM grids.

\subsection{1.2 ADF-STEM acquisition}
Atomic-resolution ADF-STEM images of the \WSeTe\ sample were acquired with an aberration-corrected Thermo Fisher Themis Z STEM, operated at 80 kV with probe current around 30 pA. The point resolution is about 1 \Ang with 25 mrad convergence semi angle. The collection semi angles were chosen to range between 63 to 200 mrad to optimize the Z-contrast and signal-to-noise ratio. To minimize the potential distortions from sample drift, we acquired 10 sequential frames with short dwell time (2$\mu$s per pixel and 20.74 pm pixel size) then perform frame-averaging using rigid registration\cite{Savitzky2018}. The sample drift rate is measured to be 5 pm/sec. The total dose for the summed image is $1.24 \times 10^7$ \dose. This dose is within the typical range for imaging 2D materials with atomic-resolution TEM and STEM of $10^5 - 10^9$ \dose \cite{Wang2016a, Algara-Siller2013, Lehnert2017, Elibol2018, Huang2013a, Robertson2016}.

\subsection{1.3 Impact of electron dose on atomic defects}
 Movie S1 is a series of sequential STEM images of \WSeTe\ showing the impact of electron dose on the atomic structure. We observe the generation, migration, and annihilation of defects as a function of electron dose. As the total dose increases, the total defect number increases. Local holes start coalescing and form larger holes after the frame 80, which corresponds to a total dose of $3.1 \times 10^8$ \dose. This suggests that the conventional approach for precision enhancement by accumulating electron dose would not work in the case of beam sensitive 2D materials. The movie contains 120 frames. Each frame was acquired at 80kV acceleration voltage with 35 pA probe current. The dwell time per pixel is $4\mu$s and the field of view is 15.3 nm $\times$ 15.3 nm. The dose for a single frame is 3.9 $\times 10^6$ \dose. Movie is sped up by 625x and downsampled to reduce the file size. 

\newpage

\section{2. Deep learning model}
We employ a fully convolutional network (FCN) with the structure of a ResUNet \cite{Zhang2018}. This is an encoder-decoder type architecture with skip connections and concatenation between each downsampling-upsampling pair. The skip connections correspond to the ResNet scheme, while concatenating the encoding blocks to decoding blocks makes up the U-Net scheme. We employ batch normalization after each convolution, and add dropout layers between each block with a dropout rate of 0.1. The model is implemented using Keras (version 2.2.4). Using Google Colab as computing nodes, we trained separate neural network for each defect type (4 in total) using the Adam optimizer scheme with a categorical cross entropy loss function. Each defect model has training and validation set with 14600 and 1624 images respectively. See section 2.2 for the details about generation of the training set. The batch size is 32, while the training epochs ranged from about 550 to 2500 and took about 12 hours to train all the models in parallel.

\subsection{2.1 FCN architecture}
\begin{figure}[H]
  \includegraphics[width=16cm]{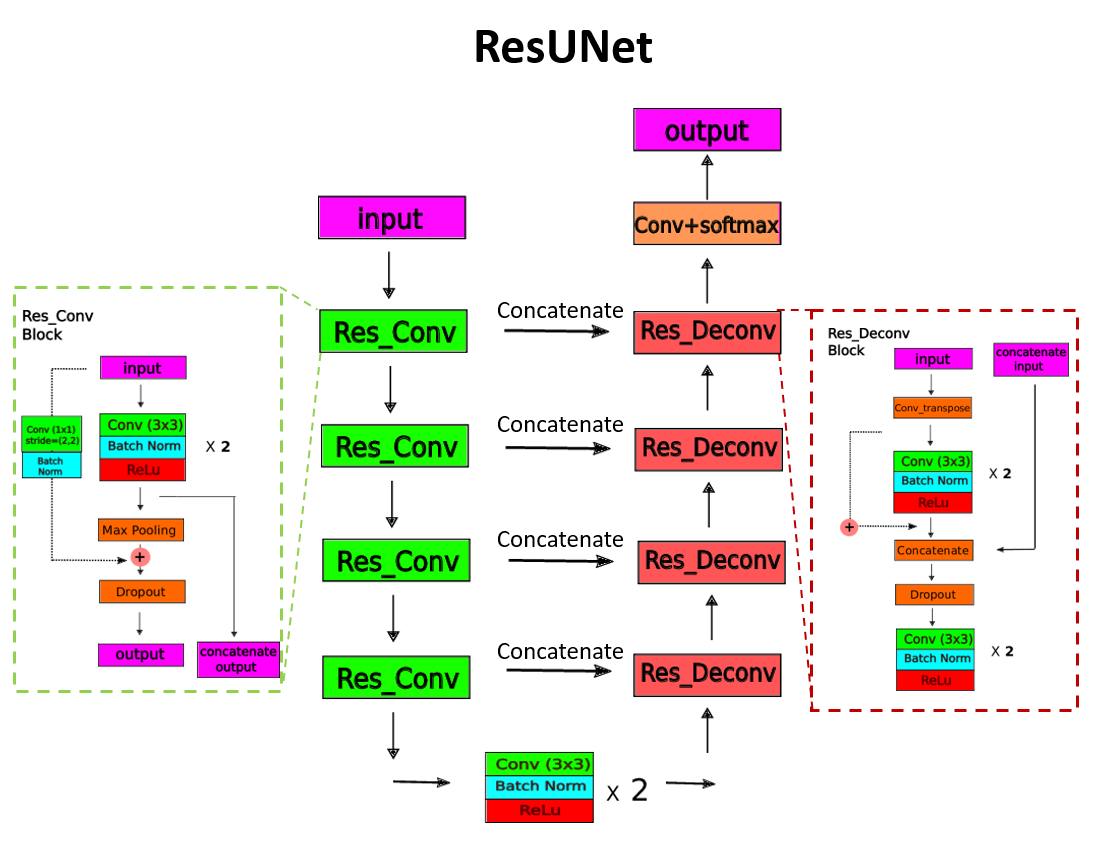}
  \caption{Illustration of the ResUNet structure. Each convolution layer was modified with skip connection inspired by ResNet, while concatenating the convolution layer with corresponding decovolution layer represents the U-Net scheme. This ResUNet structure allows us to train efficiently with pixel-wise image segmentation, which combines the advantages of ResNet and U-Net.} 
  \label{fgr:S1}
\end{figure}

We trained 4 separate FCN models for each defect type, including double vacancy ($DV$), single vacancy ($SV$), single Te substitution ($SeTe$), and double Te substitution ($2Te$). The FCN models were constructed using the ResUNet scheme with 4 res-convolution block layers and 4 res-upsampling layers as shown in Figure \ref{fgr:S1}1.

\newpage

\subsection{2.2 Generating simulated STEM images for the training set}
We simulated 50  STEM images (1024 $\times$ 1024 pixels each) as training data, shown in Figure S2. Each data set includes corresponding labels indicating the defect sites. Simulations were conducted using the Computem software using an  incoherent imaging approximation, which gives semi-quantitative results for imaging 2D materials with much faster speed than a full multislice approach. 

We wrote a script to randomly create defects at chalcogen sites and generate parameters of microscope conditions for simulations based on Gaussian distributions as shown in Table S1. The as-calculated image only contained Poisson noise. In order to make our simulations more realistic, we apply a set of post-processing steps that includes the addition of Gaussian noise, probe jittering, image shear, and background contamination extracted from the experimental images. These post-processing steps allowed us to generate very realistic training data that not only matches our experimental data, but also provides a wide range of image conditions that prevents the models from overfitting. The post-processed images were then augmented and segmented into 256 $\times$ 256 pixel windows before the model training as shown in Figure S2. The augmentation process includes rotation, flipping, and 2 types of consecutive down- and up-sampling (image size (px): 1024$\rightarrow$512$\rightarrow$1024 and 1024$\rightarrow$256$\rightarrow$1024). Figure \ref{fgr:S3}a shows the model performance including training and validation losses versus the number of epochs during the training process.  

\begin{figure}[H]
  \includegraphics[width = 16 cm]{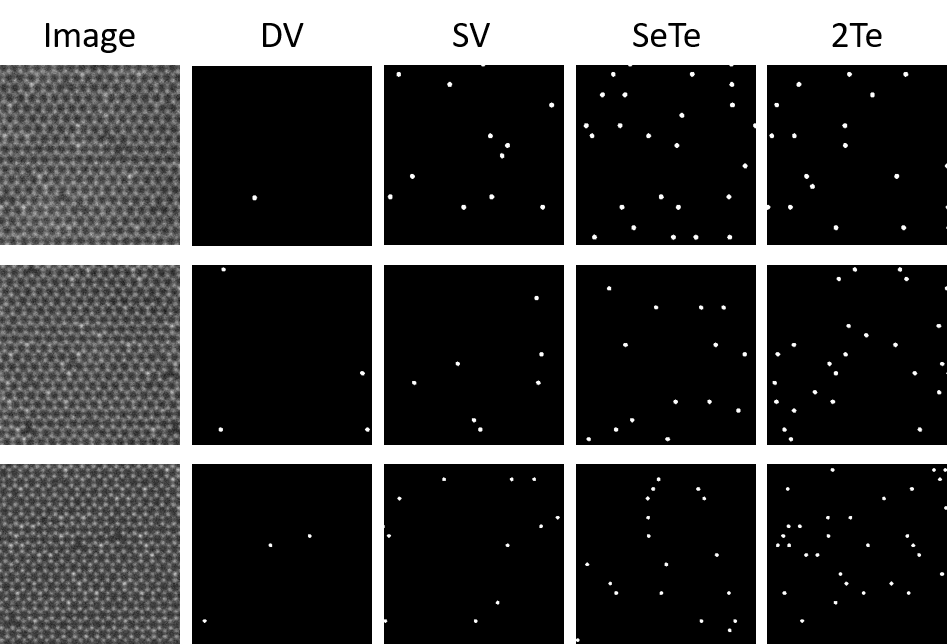}
  \caption{Simulated ADF-STEM images and defect labels. The images are calculated using Computem and then corrupted with post-processing such as Gaussian noise, probe jittering, shearing, and non-uniform background. This figure shows a small subsection of the training data: 3 images from the 50 simulated images, with a small 256 $\times$ 256 pixels window of each image for clarity. Each image is accompanied with 4 defect labels that correspond to the positions of each defect type in the image.} 
  \label{fgr:S2}
\end{figure}

\begin{table}
\caption{Simulation and post-processing parameters for training data generation}
\label{tbl:1}
\begin{tabular}{@{}cccc@{}}
\toprule
\textbf{C3 (nm)}                 & \textbf{C5 (mm)}             & \textbf{Source Size (\AA)} & \textbf{Defocus Spread (\AA)} \\ \midrule
0$\pm$ 500                       & 0$\pm$ 0.5                   & 1.05 $\pm$ 0.1             & 20 $\pm$ 10                   \\ \midrule
\textbf{Horiz. Shear (\%)}       & \textbf{Vertical Shear (\%)} & \textbf{jittering (px)}    &                               \\ \midrule
$\pm$ 5 $\pm$ 2.5                & $\pm$ 5 $\pm$ 2.5            & 0 $\pm$ 1                  &                               \\ \bottomrule
\end{tabular}
\end{table}
\newpage

\subsection{2.3 Performance evaluation of the FCN models}
The performance of the trained models were evaluated by traditional metrics in classification scenario such as precision, recall, balanced accuracy, and F1 score\cite{Murphy2012}. Note that the precision mentioned here is different from the precision for W-W separation measurement in the main text. The precision in classifications refers to how accurate the returned results from a model are, while the precision in measurement relates to how consistent the measurements are. These metrics can be calculated using the confusion matrix, which contains true-positive ($TP$), false-positive ($FP$), true-negative ($TN$), and false-negative ($FN$). The positive class denotes the defects, while the negative class denotes the non-defect sites. For example, while calculating the metrics for $2Te$ model, the $TN$ class corresponds to every chalcogen sites except $2Te$, which includes $DV$, $SV$, $SeTe$, and $2Se$. The definitions of these evaluation metrics are shown below:

\begin{equation}
    Precision = \frac{TP}{TP+FP}
\end{equation} 

\begin{equation}
    Recall \; \mbox{\textit{or True positive rate (TPR)}} = \frac{TP}{TP+FN}
\end{equation} 

\begin{equation}
    \mbox{\textit{True negative rate (TNR)}} = \frac{TN}{TN+FP}
\end{equation} 

\begin{equation}
    Balanced \; Accuracy = \frac{TPR+TNR}{2}
\end{equation} 

\begin{equation}
    F1 \; Score = 2 \cdot \frac{Precision \times Recall}{Precision + Recall}
\end{equation} 

\begin{figure}[H]
  \includegraphics[width=13cm]{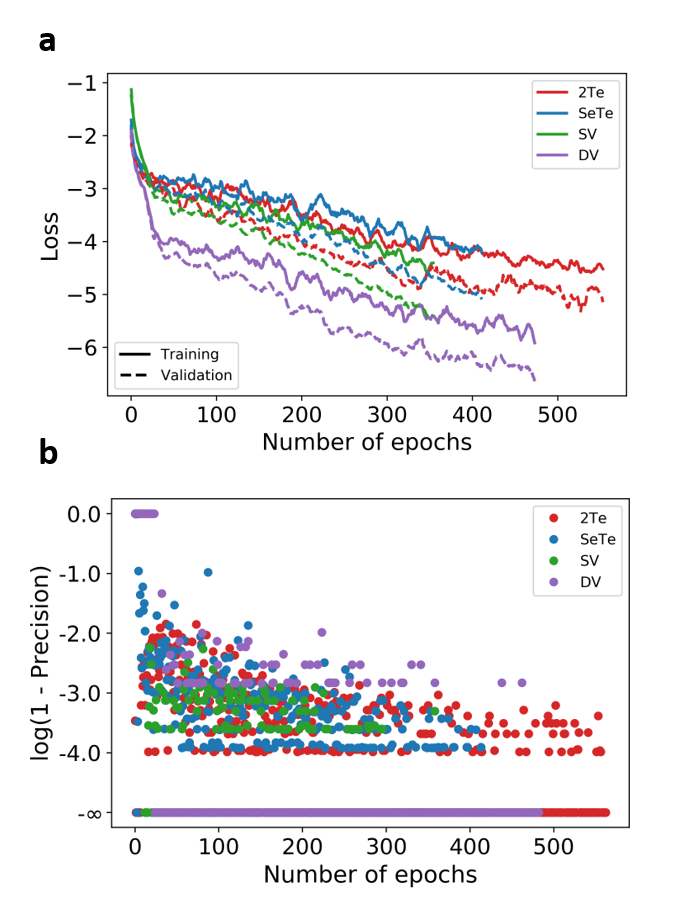}
  \caption{FCN model performance versus number of epochs. (a) Training and validation losses versus number of epochs for each defect type. (b) Plotting log(1 - $Precision$) of the validation set versus the number of epochs for each defect type. Note that the negative infinity labeled on the y-axis corresponds to a precision equal to 1.}
  \label{fgr:S3}
\end{figure}

Precision, recall, balanced accuracy, and F1 score for each defect model were calculated and shown in Table \ref{tbl:2} using simulated STEM images as test set. Computing these values on a test set that was not used for training, we find that the models show almost 99\% precision for all defect types. We compared the model predictions and the original labels to calculate the metrics. For the purpose of summing identical objects in class averaging, the most important metric is the precision of the model, which determines the ``correctness'' of the identified defects, or equivalently the ``purity'' of the defect image stack. False-positive defect images are detrimental to the overall measurement of defect structures. For example, a stack consists of 80\% $SV$ and 20\% $DV$ will skew the measured W-W separation in the class-averaged image and introduce extra spread to the precision as well than a pure stack of $SV$ while bootstrapping (see section 3.1 for discussion of bootstrapping). Note that the commonly-used pixel-wise accuracy was not used for performance evaluation since the large amount of true-negative samples (non-defect sites) will dominate the calculation, hence the model can produce an accuracy higher than 95\% by simply predicting every pixel as a non-defect site. It is worth mentioning that the units of $TP$, $FP$, $TN$, and $FN$ are chosen to be ``counts'' instead of ``pixels''. Due to the alignment procedure, the identified positions are allowed to have slight offsets to the real positions and have no effect to the following class averaging. Hence we defined a ``hit'' as long as the identified positions are within a certain distance ($\approx$150 pm) to the real positions. This distance is roughly the planar W-Se distance. Figure \ref{fgr:S3}b shows the log(1- $Precision$) versus the number of epochs. 

\begin{table}
\caption{Performance metrics of FCNs}
\label{tbl:2}
\begin{tabular}{@{}ccccc@{}}
\toprule
\textbf{Models}    & \textbf{Precision} & \textbf{Recall} & \textbf{Balanced Accuracy} & \textbf{F1 Score} \\ \midrule
DV   & 0.9955         & 0.9978      & 0.9989                 & 0.9966        \\ \midrule
SV   & 0.9989         & 0.9956      & 0.9978                 & 0.9972        \\ \midrule
SeTe & 0.9881         & 0.9781      & 0.9887                 & 0.9830        \\ \midrule
2Te  & 0.9935         & 0.9959      & 0.9977                 & 0.9947        \\ \bottomrule
\end{tabular}
\end{table}

\newpage
\section{3. Image analysis}
\subsection{3.1 Bootstrapping process}
\begin{figure}[H]
  \includegraphics[width=16cm]{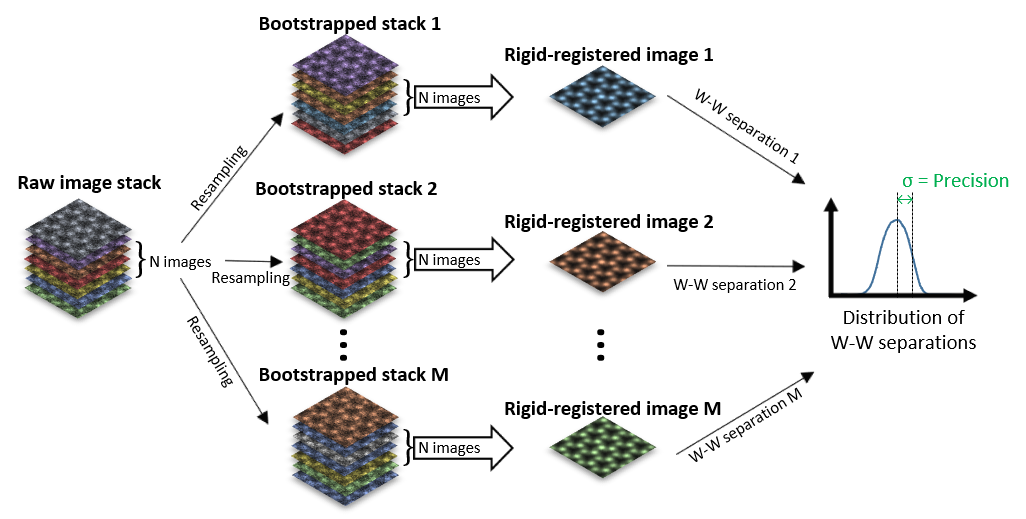}
  \caption{Illustration of bootstrapping process for precision measurement. Bootstrapped stacks were generated by resampling with replacement from the raw image stack, hence each bootstrapped stack is slightly different from the others to approximate the sampling distribution. Thus we can generate multiple rigid-registration images by bootstrapping, which allow us to measure the distribution of W-W separations and its standard deviation as measurement precision.}
  \label{fgr:S4}
\end{figure}

We applied a bootstrapping method to calculate the precision of our measurements of W-W separations of rigid-registered defect images.  This method is illustrated in Figure \ref{fgr:S4}. The goal of bootstrapping is to get the sampling distribution of an estimator from limited samples by creating resamples using a simple approach---resampling with replacement. Our bootstrapping approach works as follows. The raw image stack contains $N$ images, which is our sample size. We then randomly draw 1 image from the raw image stack for $N$ times with replacement. Here, drawing with replacement means that if an image is drawn, it is placed back in to the original stack and can be selected again. We repeat this process N times, producing a new image stack that is ``resampled'': it has the same size ($N$) as the original image stack. Note that some of the images might appear multiple times in the new image stack, and some may not show up at all. 
We repeat this process $M$ times, which gave us $M$ bootstrapped samples that were used to estimate the sampling distribution of the estimator (W-W separations in our case). For each bootstrapped sample, we generated one rigid-registered image and measured the W-W separation. The precision was then defined by the standard deviation of $M$ measurements of W-W separations as shown in Figure 3b and 3d of the main text. In our analysis, we picked $M=22500$ for all the defect types, which is much higher than the general suggestion of $M>N\ln{N}$ for the estimator of bootstrapping distribution to converge to the estimator of sampling distribution\cite{Singh2008}. We also apply this bootstrapping approach to estimate the uncertainties in our experimental strain line profile in Figure 5j of the main text.

\newpage

\subsection{3.2 Measuring precision and SNR}

\begin{figure}[H]
  \includegraphics[width=16cm]{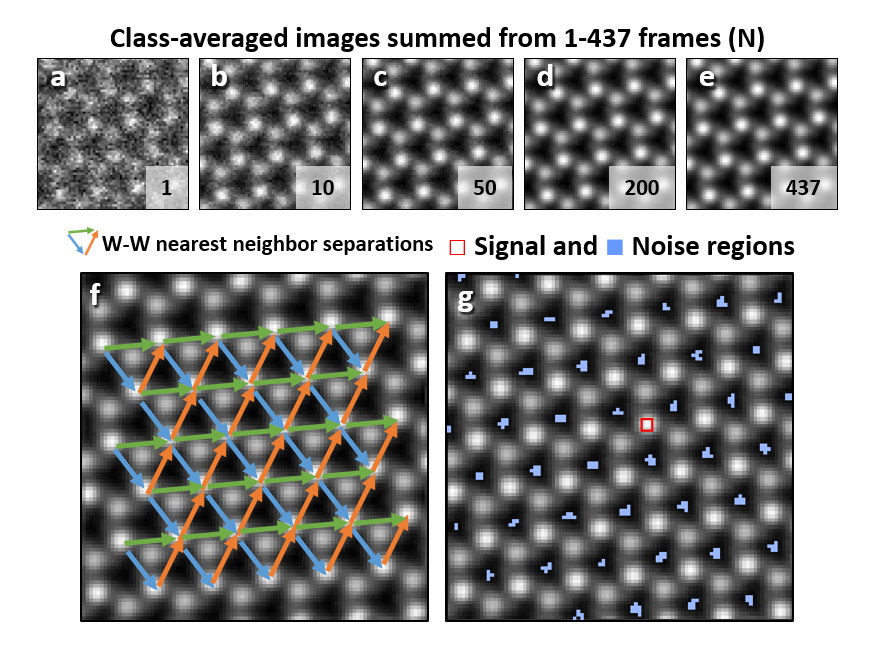}
  \caption{Measuring precision and SNR for each class-averaged image that was summed from 1-437 frames. (a-e) Class-averaged images that were summed from 1-437 frames, showing the increase of SNR as a function of summed frames. (f) W-W nearest neighbor separations, (g) assigned signal and noise regions overlaid on class-averaged image of defect-free $2Se$ sites.}
  \label{fgr:S5}
\end{figure}

This section describes the methods used to evaluate the precision and SNR gain as a function of the number of frames summed for Figure 3c. Class-averaged images were generated using a varying number of summed frames from 1-437 as shown in \ref{fgr:S5}a-e. The precision for each class-averaged image is measured by comparing the standard deviation of all W-W nearest neighbor separations in a class-averaged image of defect-free regions of \WSeTe as illustrated in \ref{fgr:S5}f. In a defect-free lattice, all W-W nearest neighbor distances should be equal; as a result, the spread in measured atomic spacings should indicate the precision of the measurement.

For measuring SNR, we defined the signal intensity as the mean intensity of a 3 px $\times $3 px window sitting on the center of W atom ($\mu_{signal}$). We defined the noise as the standard deviation of pixel intensities at certain ``background regions'' ($\sigma_{noise}$). In \ref{fgr:S5}g, the background regions are determined by selecting pixels with lowest intensities (2.5\% percentile) in the class-averaged image, which are almost always located in the empty space enclosed by hexagonal rings. The SNR is then calculated by dividing the signal intensity by standard deviation of the noise from these background regions. We found that these measurements were highly sensitive to our definition of the background region, and that defining a suitable region to reproducibly measure the  background noise is critical for reliable SNR calculation. 

\begin{equation}
    SNR = \frac{\mu_{signal}}{\sigma_{noise}}
\end{equation}

\begin{equation}
    SNR\;Gain= \frac{SNR_{summed}}{SNR_{raw}}
\end{equation}

\newpage

\subsection{3.3 Displacement field and strain analysis}
\begin{figure}[H]
  \includegraphics{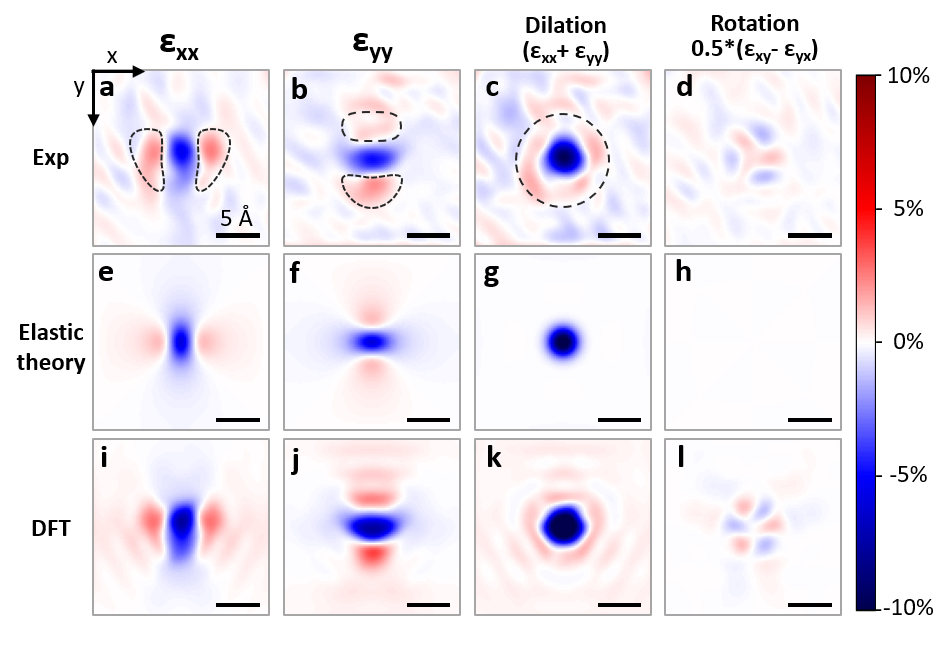}
  \caption{Strain and rotation fields at single Se vacancy site. Reprinted from Figure 5 and included rotation fields for completeness.} 
  \label{fgr:S6}
\end{figure}

The continuous displacement fields were obtained from interpolating the x and y components of the displacement vectors using the \textbf{griddata} function with \textit{cubic} option from \textbf{Scipy} package in Python. The obtained displacement fields ($D_x$ and $D_y$) were then smoothed by roughly one-third of the in-plane W-Se nearest-neighbor spacing ($\approx$ 60 pm). The 2D strain tensor components were calculated by taking spatial derivatives of the continuous displacement fields ($\epsilon_{ij} = \partial{D_i}/\partial{x_j}, {x_j} = \hat x$ or $\hat y$) as shown in Figure \ref{fgr:S6}. 

\newpage
\section{4. Density functional theory calculations}

Density functional theory \cite{Hohenberg1964, Kohn1965} (DFT) calculations were performed using the Vienna Ab-Initio Simulation Package (VASP) package \cite{Kresse1996, Kresse1999}.
We used the generalized-gradient approximation by Perdew, Burke, and Ernzerhof (PBE) to describe exchange and correlation \cite{Perdew1996}.
Kohn-Sham states were expanded into a plane-wave basis  up to a cutoff energy of 400 eV.
The electron-ion interaction was described using the projector-augmented wave method \cite{Blochl1994}.
We studied pristine \ce{WSe2} and a single Se vacancy defect using a $9 \times 9$ super cell, in which periodic images of the \ce{WSe2} materials were separated by 17 \Ang of vacuum along $z$ direction. 
The corresponding Brillouin Zone was sampled using a $1 \times 1 \times 2$ Monkhorst-Pack $\mathbf{k}$-point grid \cite{Monkhorst1976}.
Cell shape and all atomic positions were relaxed until all Hellman-Feynman forces were smaller than 0.1 meV/\Ang.
We found the optimized lattice parameter of pristine \ce{WSe2} to be 3.316 \Ang, in agreement with a previously reported value\cite{Le2015}.

\bibliography{SupInfo}